# VIPERS: An Unprecedented View of Galaxies and Large-scale Structure Halfway Back in the Life of the Universe

Luigi Guzzo[1]
and the VIPERS Team*

[1] INAF, Osservatorio di Brera, Milano, Italy

The VIMOS Public Extragalactic Redshift Survey (VIPERS) is an ongoing Large Programme to map in detail the large-scale distribution of galaxies at 0.5 < $z$ < 1.2. With a combination of volume and sampling density that is unique for these redshifts, it focuses on measuring galaxy clustering and related cosmological quantities as part of the grand challenge of understanding the origin of cosmic acceleration. Moreover, VIPERS has been designed to guarantee a broader legacy, allowing detailed investigations of the properties and evolutionary trends of $z \sim 1$ galaxies. The survey strategy exploits the specific advantages of VIMOS, aiming at a final sample of nearly 100 000 galaxy redshifts to $i_{AB}$ = 22.5 mag, which represents the largest redshift survey ever performed with ESO telescopes. In this article we describe the survey construction, together with early results based on a first sample of 55 000 galaxies.

* The VIPERS Team:
L. Guzzo[1] (P.I.), U. Abbas[2], C. Adami[3], S. Arnouts[4],
J. Bel[5], M. Bolzonella[6], D. Bottini[7], E. Branchini[8],
A. Burden[9], A. Cappi[6, 10], J. Coupon[11], O. Cucciati[6],
I. Davidzon[12, 6], S. de la Torre[13], G. De Lucia[14],
C. Di Porto[6], P. Franzetti[7], A. Fritz[7], M. Fumana[7],
B. Garilli[7], B. R. Granett[1], L. Guennou[3], O. Ilbert[3],
A. Iovino[1], J. Krywult[15], V. Le Brun[3], O. Le Fevre[3],
D. Maccagni[7], K. Malek[16], A. Marchetti[17, 1],
C. Marinoni[5], F. Marulli[11], H. J. McCracken[18],
Y. Mellier[18], L. Moscardini[11], R. C. Nichol[9], L. Paioro[7],
J. A. Peacock[12], W. J. Percival[9], S. Phleps[19],
M. Polletta[7], A. Pollo[20, 21], H. Schlagenhaufer[19],
M. Scodeggio[7], A. Solarz[16], L. Tasca[3], R. Tojeiro[9],
D. Vergani[22], M. Wolk[18], G. Zamorani[6], A. Zanichelli[23]

[1] INAF, Osservatorio di Brera, Milano, Italy; [2] INAF Osservatorio di Torino, Italy; [3] LAM, Marseille, France; [4] Canada-France-Hawaii Telescope, Hawaii, USA; [5] CPT Universite de Provence, Marseille, France; [6] INAF, Osservatorio di Bologna, Italy; [7] INAF, IASF Milano, Italy; [8] Universitá Roma 3, Rome, Italy; [9] ICG, University of Portsmouth, United Kingdom; [10] Université de Nice, Obs. de la Cote d'Azur, Nice, France; [11] Inst. of Astron. and Astrophys., Academia Sinica, Taipei, Taiwan; [12] Dip. di Fisica e Astronomia, Universitá di Bologna, Italy; [13] Institute for Astronomy, University of Edinburgh, United Kingdom; [14] INAF, Osservatorio di Trieste, Italy; [15] Jan Kochanowski University, Kielce, Poland; [16] Dept. of Particle and Astrophys. Science, Nagoya Univ., Japan; [17] Universitá degli Studi di Milano, Italy; [18] IAP, Paris, France; [19] MPE, Garching, Germany; [20] Astronomical Observ., Jagiellonian University, Cracow, Poland; [21] National Centre for Nuclear Research, Warsaw, Poland; [22] INAF, IASF Bologna, Italy; [23] INAF, IRA Bologna, Italy

There is no doubt that one of the major achievements of observational cosmology in the 20th century has been the detailed reconstruction of the large-scale structure of the Universe around us. Starting in the 1970s these studies developed into what nowadays is the industry of redshift surveys, beautifully exemplified by the ever-growing Sloan Digital Sky Survey project (SDSS, e.g., Eisenstein et al., 2011).

Maps of the large-scale galaxy distribution have shown not only that the topology of large-scale structure is quite different from how it was imagined at the time of Edwin Hubble and Fritz Zwicky, but have also been crucial to quantitatively supporting the current, successful model of cosmology. The inhomogeneity that we can measure in the galaxy distribution on different scales is one of the most important relics of the initial conditions that shaped our Universe. The observed shape of the power spectrum $P(k)$ of density fluctuations (or its Fourier transform, the correlation function $\xi(r)$), indicates that we live in a Universe in which only ~ 25% of the mass–energy density is provided by (mostly dark) matter. Combined with other observations, it also implies that a ubiquitous repulsive "dark energy" is required to provide the remaining ~ 75% and make sense of the overall picture.

The peculiar motions of galaxies, which reflect the overall growth of structure driven by gravitational instability, also produce measurable effects on these clustering measurements. They provide a way to check whether the "dark energy" hypothesis is really correct, or rather that the observed acceleration is indicating a more radical possibility, i.e., that the theory of general relativity (GR), describing the force of gravity, needs to be revised on large scales.

Galaxy velocities affect the measured redshifts and produce what are known as redshift space distortions (RSD) in the maps of large-scale structure (Kaiser, 1987; Peacock et al., 2001) and, in turn, in the measured galaxy correlations. The observed anisotropy is proportional to the growth rate of cosmic structure $f(z)$, which is a trademark of the gravity theory: if GR holds, we expect to measure a growth rate $f(z) = [\Omega_m(z)]^{0.55}$ (Peebles, 1980). RSD are now recognised as one of the primary ways to make this test (Guzzo et al., 2008). Directly probing the amplitude and anisotropy of clustering, redshift surveys promise to play a major role also in 21st-century cosmology, at least as important as they did in the past one, as several planned experiments, including the forthcoming ESA Euclid mission (Laureijs et al., 2011), testify.

However, the yield of a redshift survey is much more than this. By building statistically complete samples of galaxies with measured luminosity, spectral properties and often colours and stellar masses, they are a key probe of galaxy formation and evolution and of the relationship between the baryonic component that we observe and the hosting dark-matter halos. A survey like the SDSS, for example, based on one million redshifts, was able to measure to exquisite precision global galaxy population trends involving properties such as luminosities, stellar masses, colours and structural parameters (e.g., Kauffmann et al., 2003).

In more recent years, deeper redshift surveys over areas of 1–2 square degrees have focused on exploring how this detailed picture emerged from the distant past. Most of these efforts saw the VLT and the VIMOS spectrograph play a central role, specifically in the case of the VVDS (VIMOS–VLT Deep Survey; Le Fevre et al., 2005) and zCOSMOS (Lilly et al., 2009) surveys. The main goal of these projects was to trace galaxy evolution back to its earliest phases and/or understand its relationship with environment. Only the Wide extension of the VVDS (Garilli et al., 2008), encompassed sufficient volume to attempt cosmologically meaningful computations (Guzzo et al., 2008), but with large error bars. At the end of the past decade it therefore became clear to us that a new step in deep redshift surveys was needed, in the direction of building a sample at $z \sim 1$ with volume and statistics comparable to those of the available surveys of the local Universe.

## VIPERS

VIPERS was conceived to fill this gap by exploiting the unique capabilities of VIMOS. Started in Period 82, the survey aimed to measure redshifts for ~ $10^5$ galaxies at $0.5 < z < 1.2$, covering an unprecedented volume. Its goals are to accurately and robustly measure galaxy clustering, the growth of structure and galaxy properties at an epoch when the Universe was about half its current age. The galaxy target sample is based on the excellent five-band photometric data of the Canada–France–Hawaii Telescope Legacy Survey Wide catalogue (CFHTLS–Wide[1]).

To achieve its goals, VIPERS covers ~ 24 square degrees with a mosaic of 288 VIMOS pointings, split over two areas in the W1 and W4 CFHTLS fields (192 and 96 pointings respectively). With a total exposure time of 45 minutes per VIMOS pointing, VIPERS will use a total of 372 hours of multi-object spectrograph (MOS) observations, plus 68.5 hours of pre-imaging. Its area and depth correspond to a volume of $5 \times 10^7$ $h^{-3}$ $Mpc^3$, similar to that of the 2dF Galaxy Redshift Survey at $z \sim 0$ (Colless et al., 2001). Such a combination of sampling and volume is unique among redshift surveys at $z > 0.5$. The target sample includes all galaxies with $i_{AB} < 22.5$ mag, limited to $z > 0.5$ through a robust $ugri$ colour pre-selection. The use of an aggressive short-slit strategy (Scodeggio et al., 2009), allows the survey to achieve a sampling rate > 40%. The VIMOS low-resolution red grism (spectral resolution $R = 210$), yields a spectral coverage between 5500 and 9500 Å, for a typical redshift root mean square (rms) error of $\sigma_z = 0.00047$ $(1 + z)$ (directly estimated from double measurements of 1215 galaxies). These and more details on the survey construction and the properties of the sample can be found in Guzzo et al. (2013).

## Survey and data management

Handling a spectroscopic survey of this size requires the automation of most of the typical operations and that human intervention is reduced to the minimum necessary. Building on the experience accumulated with previous surveys using VIMOS, a fully automated pipeline to process the observations from the raw frames to the final redshift measurement was set up at INAF–IASF in Milan. In this scheme, human work is limited only to the final verification and validation of the measurements, which is still necessary to recover about 20% of the correct redshifts for the lowest quality data.

Most importantly, the whole management and book-keeping of the survey process was also organised within a web-based environment, named EasyLife (Garilli et al., 2012). This is a key feature of VIPERS, which allows us to control in real time any past, present and future events concerning each of the 288 VIMOS pointings that compose the VIPERS mosaic, such as pre-imaging, mask preparation, real-time atmospheric conditions during the MOS observations, quality of the spectra, names of the redshift reviewers, etc. This is achieved through a point-and-click graphical interface, whose appearance is shown in Figure 1. The two panels give the real-time status of each VIMOS pointing in the two W1 and W4 VIPERS fields.

Similarly, the complete work output of the team, from data validation to the science analyses, the distribution of results and draft papers, can be monitored and co-ordinated through related web pages. Finally, validated redshifts and spectra are fed into an SQL-based database at the completion of the validation phase, such that the real-time survey catalogue (visible only to the data handling group) is constantly and automatically updated to the very last redshift measured. Internal releases can then be made by simply creating a snapshot of the catalogue at a given moment, assuring full version control on the catalogues used for science investigations (internally or publicly) at any stage of the project. A dedicated Wiki-page system provides the team with the appropriate environment for science discussions and to share results. Public web pages are also available, to provide updates about the general project progress[2].

## The VIPERS PDR-1 catalogue

After processing, reduction and validation, the data observed up until January 2012 yielded a catalogue of 55 358 redshifts, representing about 60% of the final expected sample. This catalogue, corresponding to the red areas in

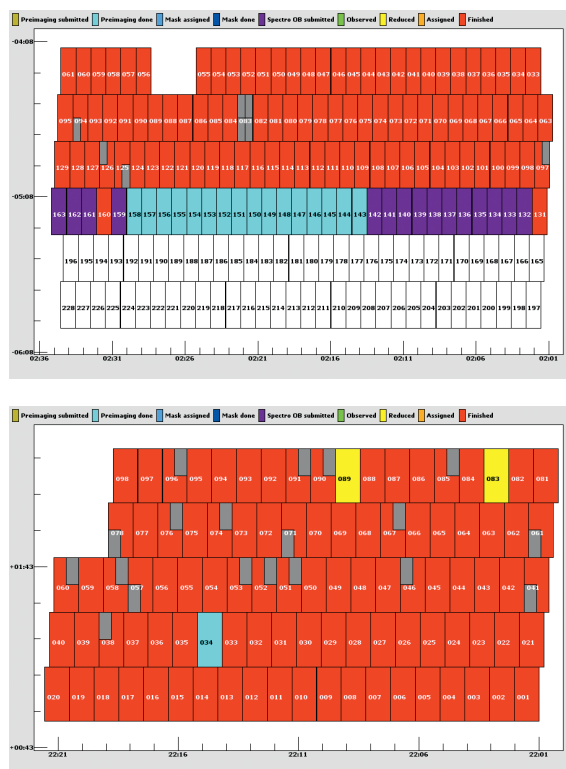

Figure 1. Snapshots from the VIPERS project monitoring web page, giving the detailed status of each VIMOS pointing that is part of the survey. The two images shown here refer to the PDR-1 catalogue discussed in the text, which contains data observed up to January 2012. These correspond to the areas depicted in red. Other colours mark different phases of the data acquisition/processing. The large number of grey rectangles in W4 (bottom) represents VIMOS quadrants for which the mask insertion failed. As can be seen, field W4 is nearly complete.

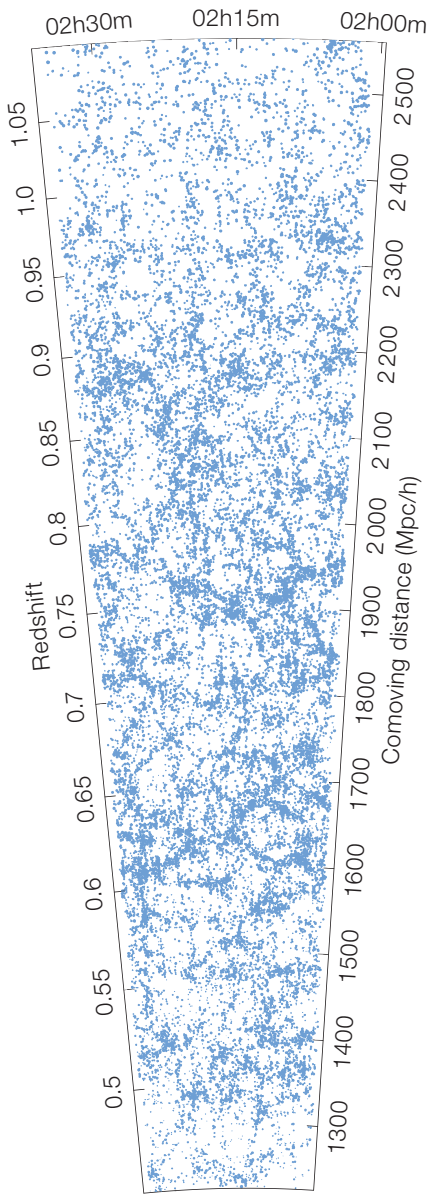

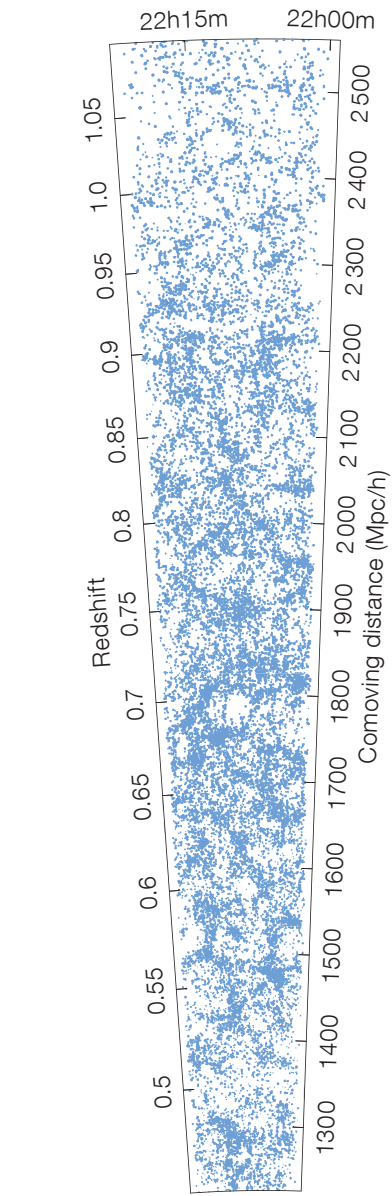

Figure 2. The stunning view of the large-scale structure of the Universe at 0.45 < z < 1.1, provided by the distribution of ~ 25 000 galaxies in the VIPERS W1 field (Guzzo et al., 2013). The opening angle corresponds to right ascension and the data are projected over ~ 1 degree in declination. The size of each dot is proportional to the galaxy B-band luminosity.

Figure 3. Same as Figure 1, but for the W4 area. The declination projection here is ~ 1.6 degrees.

Figure 1, has been frozen and is now called the Public Data Release 1 (PDR-1) catalogue. All data and analyses presented in the journal papers recently submitted for publication are based on this catalogue, which will become fully public in September 2013. Here we present a selection of these recent results.

### Unveiling the structure of the younger Universe

The first immediate outcome of VIPERS is shown by the maps of the galaxy distribution in Figures 2 and 3. The two cone diagrams present a remarkable combination of volume and dynamical range (in terms of scales sampled), which is a unique achievement of VIPERS at these redshifts. They allow us to appreciate both the details and the grand picture of large-scale structure, when the Universe was only between five and eight billion years old.

The following step is to quantify the observed structure as a function of scale. One of the main goals of VIPERS is the measurement of the amplitude and anisotropy of the galaxy two-point correlation function: a first estimate from the PDR-1 sample is shown in the two panels of Figure 4. Crucial for these measurements is an accurate knowledge of several ancillary pieces of information from the survey, such as the photometric and spectroscopic angular selection masks, the target sampling rate and the spectroscopic success rate. These allow us to assign a weight to every observed galaxy in the survey to correct for the overall incompleteness introduced by the different factors.

The fingerprint of RSD is evident in the flattening of $\xi(r_p,p)$ along the line-of-sight direction (left panel of Figure 4). This anisotropy yields a first estimate of the mean growth rate of structure, which is presented in de la Torre et al. (2013). The measured value is in agreement with the predictions of general relativity within the current uncertainty of ~ 17%. Comparison of the projected correlation functions $w_p(r_p)$ (right panel, Figure 4) shows how well the measurements from the two independent fields agree, in both shape and amplitude. The final VIPERS catalogue will allow us to extend such measurements over two or more redshift bins. Statistical errors will also be reduced by measuring RSD with two independent galaxy populations (McDonald & Seljak, 2009), an important specific advantage allowed by the high sampling of VIPERS.

Figure 5 shows another way of using the VIPERS data and quantifying galaxy clustering. The panels show different estimates of the power spectrum $P(k)$ obtained from the de-projection of the angular clustering measured over the full photometric parent sample of VIPERS i.e. the ~ 140 square degrees of the CFHTLS (Granett et al., 2012). This has been made possible by knowledge of the galaxy redshift distribution provided by an earlier version of the VIPERS catalogue. This measurement of $P(k)$ has been used to obtain improved constraints on the number of neutrino species and their total mass (Xia et al., 2012). A direct measurement of the 3D distribution of $P(k)$ from the VIPERS redshift data alone is under development.

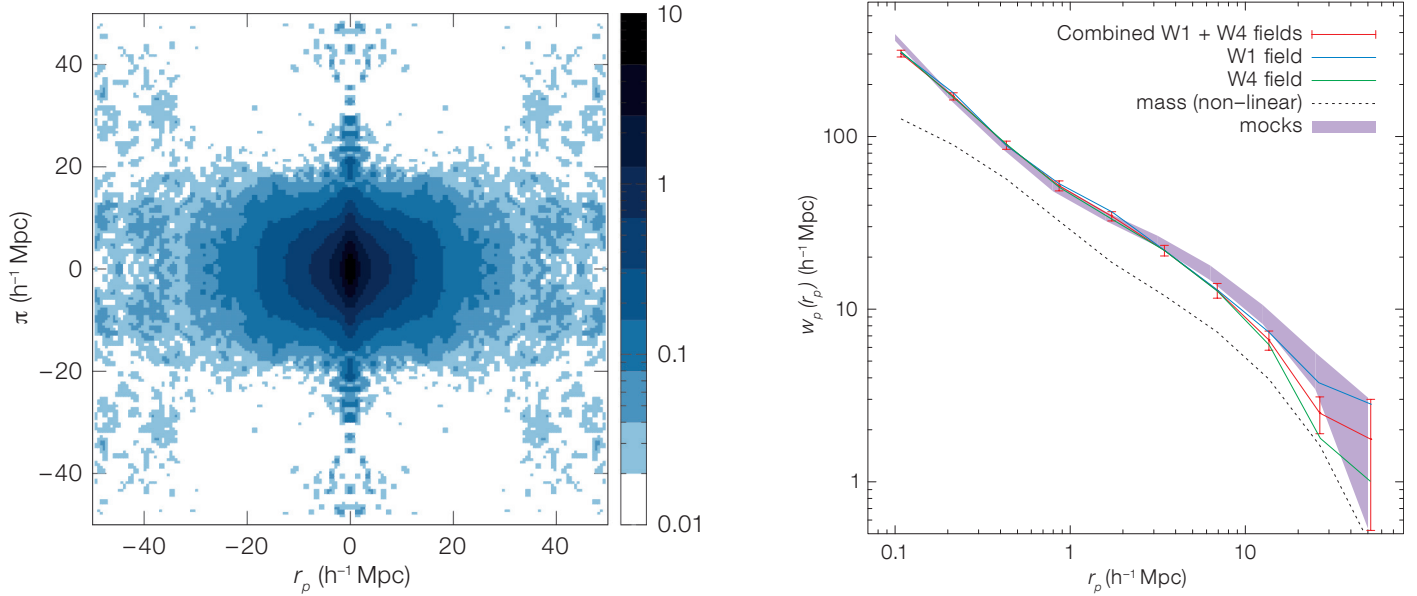

Figure 4. The redshift-space two-point correlation function over the full $0.5 < z < 1.0$ range, from the VIPERS PDR-1 catalogue. Left: The 2D correlation function $\xi(r_p,\pi)$, showing the well-defined signature of linear redshift distortions, i.e. the oval shape of the contours (de la Torre et al., 2013). Right: The projected correlation function (obtained by integrating the data in the first quadrant of the upper figure along the $\pi$ direction), for W1 and W4 fields separately and for the total sample. These are compared to the best-fitting $\Lambda$ Cold Dark Matter model for the mass (dotted line, prediction using the HALOFIT code). The shaded area corresponds to the 1$\sigma$ error corridor, computed from the scatter in the measurements of a large set of mock surveys, custom built for VIPERS.

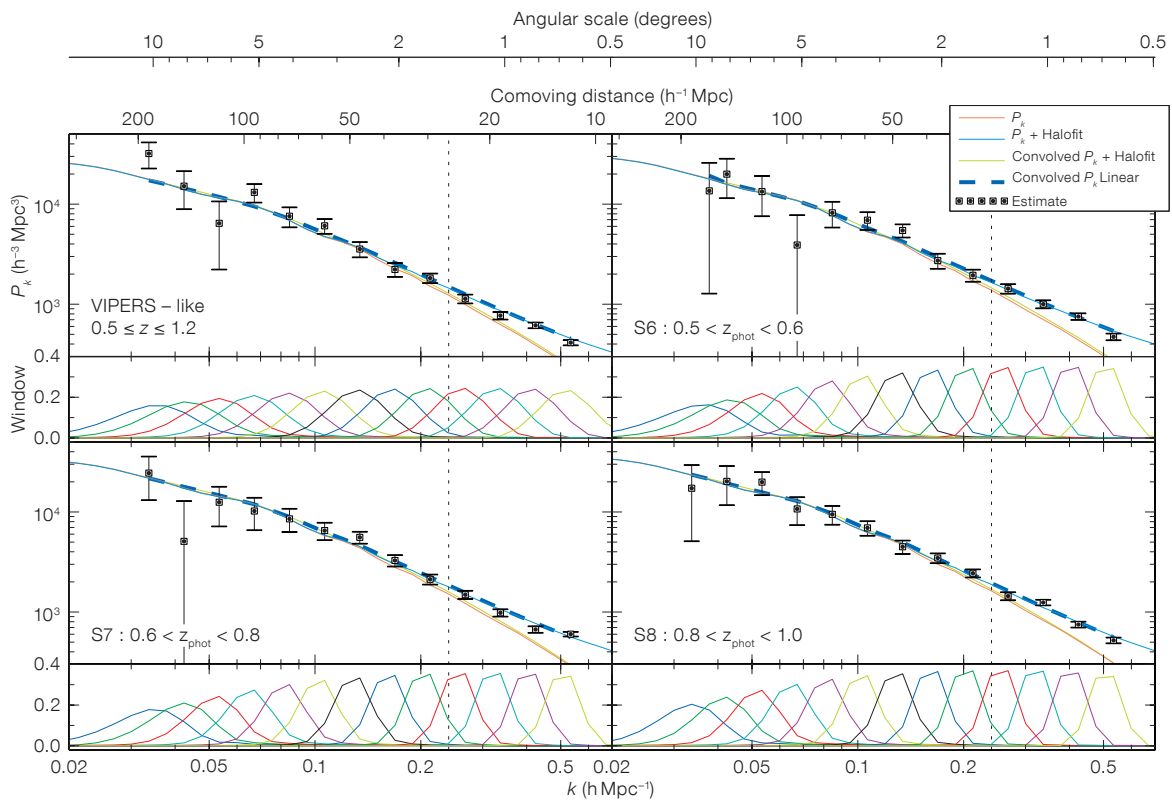

Figure 5. Early estimates of the real-space galaxy power spectrum $P(k)$, through a combination of the photometric data from the full CFHTLS area (140 square degrees) and the redshift distribution $dN/dz$ from VIPERS (Granett et al., 2012). The window function of each band is shown in the panel below each plot. The angular power spectrum is estimated from a VIPERS-like colour-selected sample (upper-left panel), and different slices in photometric redshifts are shown in the other three panels. The angular power spectrum is then de-projected to obtain the spatial power spectrum, knowing the $dN/dz$ and selection function from VIPERS.

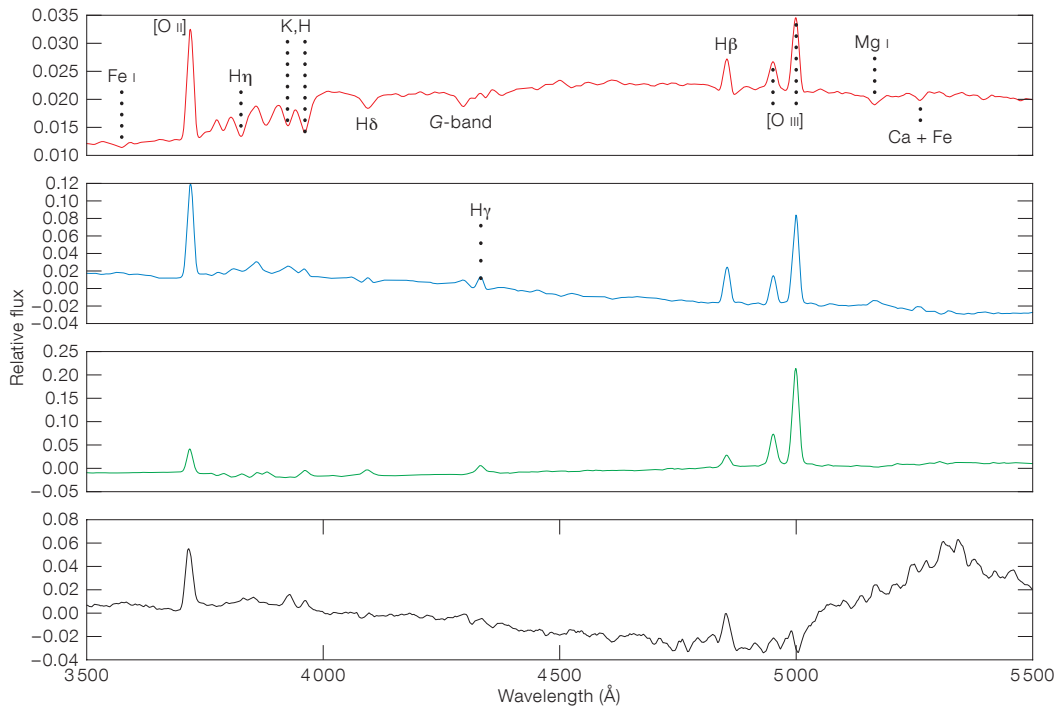

Figure 6. Plot of the first four eigenspectra obtained from the principal component analysis of about 11 000 VIPERS spectra (Marchetti et al., 2013). Each spectrum can be classified in terms of the amplitude of each of its principal components, when it is decomposed into a linear combination of these. This technique allows us to build a spectral classification scheme that is complementary to the SED-fitting technique. In contrast, specific spectra (such as those of active galactic nuclei) can also be spotted among normal galaxies as not being well reproduced by the simple combination of the principal eigenvectors.

### Pinpointing the properties of the galaxy population seven billion years ago

Figure 6 refers to another recently published result, i.e., the decomposition of the first ~ 11 000 VIPERS spectra based on principal component analysis (PCA; Marchetti et al., 2013). The main goal here has been to develop an objective classification of galaxy spectra, capable of separating different populations in a robust way. This will have applications for studies in both galaxy evolution and cosmology (e.g., to define complementary large-scale structure tracers). The application of PCA to the whole PDR-1 sample is ongoing.

As in the case of other deep surveys, an important feature of VIPERS is the complementary photometric information over a wide wavelength range. For VIPERS, the five high-quality bands of the CFHTLS have been further enriched with ultraviolet (UV; Galex) and near-infrared $K$-band (WIRCAM) photometric observations. These measurements are combined to obtain, for all VIPERS galaxies, reliable spectral energy distribution (SED) fits and, in turn, luminosities and stellar masses. Figure 7 graphically shows the unique power of combining large-scale (positions) and small-scale (here galaxy colour) information: the fact that the colour–density relation of galaxies is already in place at these redshifts is obvious from this figure. Such a wealth of information can be quantitatively encapsulated by statistical measurements of the global population, to reveal overall evolutionary trends. One such statistic is the galaxy stellar mass function; in Figure 8 we show our estimate of this from the VIPERS PDR-1 catalogue and Davidzon et al. (2013) provide more plots and details. Given the volume of VIPERS, Figure 8 presents, without doubt, the most precise estimate ever of the bright end of the mass function at $z \sim 1$ (Davidzon et al., 2013), establishing a new reference for the mean density of massive galaxies at $0.5 < z < 1.2$.

### Further results and future perspectives

There are several other important results being obtained from the current VIPERS data. These include an estimate of the matter density parameter $\Omega_m$ at $z \sim 0.8$ (Bel et al., 2013), the luminosity dependence of clustering (Marulli et al., 2013) and the non-linearity of the galaxy biasing function (Di Porto et al., 2013), the evolution of the colour–magnitude relation and luminosity function of galaxies (Fritz et al., 2013) and an automatic classification of stars, galaxies and active galactic nuclei (AGN) based on Support Vector Machines (SVM; Malek et al., 2013).

The PDR-1 catalogue includes only ~ 60 % of the final expected data. During the summer and autumn of 2012, 31 additional pointings were observed, covering another full row in W1 (the purple and cyan fields in the upper panel of Figure 1). These data are being processed and, based on the usual yield, are expected to deliver around 11 000 new redshifts. With the current pace, we expect spectroscopic observations at Paranal to be completed by 2014.

We can foresee a number of exciting results being obtained from the VIPERS data in the coming years, with many that will originate from the general community, once the PDR-1 sample is in the public domain in September 2013. The grand view of large-scale structure in the young

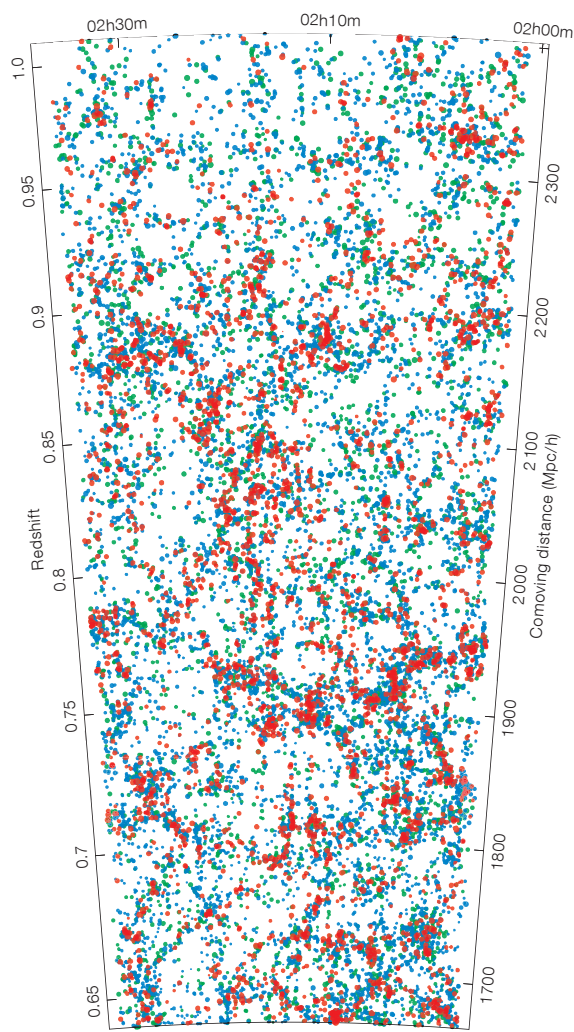

Figure 7. A zoom into the fine structure of the W1 field galaxy distribution, with an additional dimension provided by galaxy restframe colours. This picture gives an idea of the power of coupling multiband photometry to galaxy positions, when scales well above $100\,h^{-1}$ Mpc are mapped. In this example, galaxies have been marked in red, green or blue, depending on their $U$–$B$ restframe colour. In addition, the size of each dot is proportional to the $B$-band luminosity of the corresponding galaxy. The result shows, in all its glory, the colour–density relation for galaxies, already in place at these redshifts, with red early-type galaxies tracing the backbone of structure and blue/green star-forming objects filling the more peripheral lower-density regions.

Universe provided by the first data we have discussed here, makes us hope that — as has happened with previous redshift surveys — the most exciting results will be those that were not even mentioned in the original proposal, but are there awaiting us, written inside the large-scale distribution of galaxies.

Acknowledgements

We acknowledge the continuous support of the ESO staff for all operations in Garching and at Paranal. We especially thank our project support astronomer, Michael Hilker, for his enthusiastic contribution to all phases of the observation preparation and realisation chain.

Links

[1] CFHTLS: http://www.cfht.hawaii.edu/Science/CFHLS/
[2] VIPERS public web pages: http://vipers.inaf.it

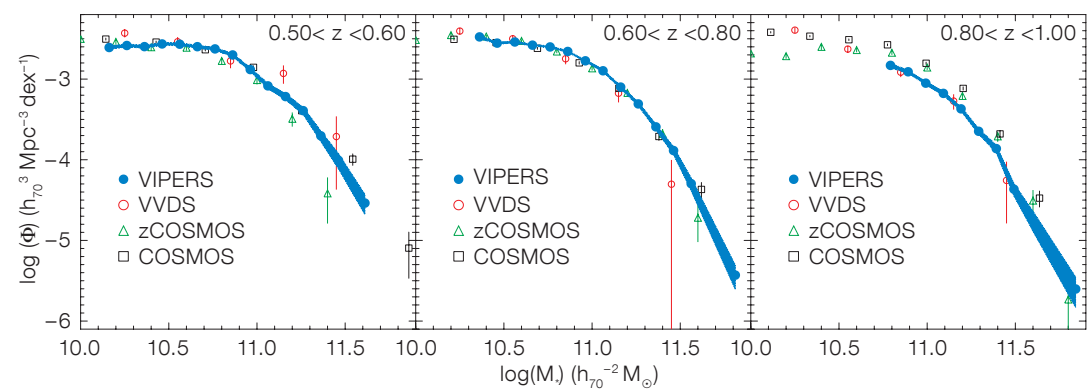

Figure 8. The VIPERS galaxy stellar mass function (MF) at three reference redshifts, estimated from the PDR-1 catalogue (blue points with error corridor, Davidzon et al., 2013). Comparison is provided with previous estimates from two spectroscopic VLT surveys (Pozzetti et al., 2007; 2010) and from the COSMOS photo-$z$ sample (Ilbert et al., 2010). Note the small size of the error corridor, indicating how, at large masses, VIPERS is providing a remarkably precise measurement for these redshifts, owing to its large volume.